\documentclass[%
 reprint,
superscriptaddress,
nofootinbib,
 amsmath,amssymb,
secnumarabic
]{revtex4-1}

\usepackage{graphicx}
\usepackage{dcolumn}
\usepackage{bm}
\usepackage{float}

\usepackage[backref=none]{hyperref} 
\usepackage[usenames,dvipsnames]{color}
\definecolor{MyDarkBlue}{rgb}{0,0.1,0.7}
\hypersetup{pdfborder={0 0 0},colorlinks,breaklinks=true,
  urlcolor={MyDarkBlue},citecolor={MyDarkBlue},linkcolor={MyDarkBlue} }

\usepackage[usenames,dvipsnames]{xcolor}

\usepackage[noend]{algpseudocode}
\usepackage{algorithm,algorithmicx}
\algrenewcommand\algorithmicrequire{\textbf{Precondition:}}
\algrenewcommand\algorithmicensure{\textbf{Postcondition:}}

\usepackage{tikz}
\usepackage{graphicx}
\usepackage{tikzscale}
\usepackage{tikz-3dplot}
\usetikzlibrary{decorations.pathreplacing}
\usetikzlibrary{fit,calc,positioning,decorations.pathreplacing,matrix}

\pgfdeclarelayer{background}
\pgfdeclarelayer{foreground}
\pgfsetlayers{background,main,foreground}

\usepackage{amsthm} 
\newtheorem{theorem}{Theorem}
\newcommand*{\etc}[1]{{#1}}

\usepackage{braket}
\DeclareRobustCommand\openone{\leavevmode\hbox{\small1\normalsize\kern-.33em1}}
\DeclareRobustCommand\openonesmall{\leavevmode\hbox{\footnotesize1\small\kern-.30em1}}

\usepackage{graphicx,bm}
\usepackage[caption=false]{subfig}

\usepackage[noend]{algpseudocode}
\usepackage{algorithm,algorithmicx}

\newcommand*\Let[2]{\State #1 $\gets$ #2}
\algrenewcommand\algorithmicrequire{\textbf{Precondition:}}
\algrenewcommand\algorithmicensure{\textbf{Postcondition:}}

\begin{document}

\title{Decoding Across the Quantum LDPC Code Landscape}

\author{Joschka Roffe}
\thanks{Contact: \href{mailto:joschka@roffe.eu}{joschka@roffe.eu}}
\affiliation{University of Sheffield}
\author{David R White}
\affiliation{University of Sheffield}
\author{Simon Burton}
\affiliation{University College London}
\author{Earl Campbell}
\affiliation{University of Sheffield}
\date{\today}

\begin{abstract}
We show that belief propagation combined with ordered statistics post-processing is a general decoder for quantum low density parity check codes constructed from the hypergraph product. To this end, we run numerical simulations of the decoder applied to three families of hypergraph product code: topological codes, fixed-rate random codes and a new class of codes that we call semi-topological codes.  Our new code families share properties of both topological and random hypergraph product codes, with a construction that allows for a finely-controlled trade-off between code threshold and stabilizer locality. Our results indicate thresholds across all three families of hypergraph product code, and provide evidence of exponential suppression in the low error regime. For the Toric code, we observe a threshold in the range $9.9\pm0.2\%$. This result improves upon previous quantum decoders based on belief propagation, and approaches the performance of the minimum weight perfect matching algorithm.  We expect \etc{semi-topological codes to have the same threshold as Toric codes, as they are identical in the bulk, and we present numerical evidence supporting this observation.}

\end{abstract}

\maketitle
\newpage

\section{Introduction}

Any scalable computer architecture must be robust against hardware imperfections. In quantum computing, where qubits are realised as fragile quantum two-level systems, fault tolerance necessitates active error correction \cite{Shor1995,Gottesman09,Devitt13,Terhal15,roffe2019quantum}. A quantum error correction \textit{code} specifies an encoding in which quantum data is distributed across a larger space of qubits to create a logical qubit state. Errors are detected on the logical state via a series of non-destructive stabilizer measurements (quantum parity checks) yielding an error \textit{syndrome}. This syndrome information is processed by a \textit{decoding} algorithm to determine the best \textit{recovery} operation to return the encoded quantum information to its uncorrupted state. All three stages of the error correction cycle --- syndrome measurement, decoding and recovery --- must be performed within a short time frame before the qubits irreversibly decohere. Performing the decoding in real-time is a computationally intensive inference problem, \etc{with realistic resource estimates showing a need for} terabytes of syndrome information to be processed per second \cite{delfosse2020}. As such, efficient decoding algorithms are necessary to allow quantum error correction to be performed whilst maintaining realistic demands on classical co-processors \cite{babar2015fifteen}.

Low density parity check (LDPC) codes are a ubiquitous classical error correction protocol \cite{gallager1962}, finding use, for example, in the recent 5G communication standard \cite{Bae2019}. The specific advantage of LDPC codes is that they can be decoded using an algorithm from probabilistic graph theory known as iterative belief propagation (BP) \cite{pearl1982reverend}. The BP algorithm exploits the structure of the error correction code to solve the decoding inference problem in time linear in the code block length \cite{kschischang2001factor}. For certain LDPC codes, BP decoding enables error correction at close to the Shannon-capacity, the theoretical upper bound on the rate of information transfer along a noisy channel \cite{shannon49,mackay1997near}.

Quantum LDPC (QLDPC) codes can be constructed from classical LDPC codes using the hypergraph product framework due to Tillich and Zemor \cite{tillich2013quantum}. The hypergraph product translates the parity check sequences of a classical \textit{parent} code into a set of commuting stabilizers that define a quantum code. \etc{The most commonly studied hypergraph product codes fall into one of two types}: \textit{topological} QLDPC codes and \textit{random} or \textit{expander} QLDPC codes.

In contrast to classical LDPC codes, there is no established decoder that works generally for all QLDPC codes. For purely 2D topological codes, the minimum weight perfect matching algorithm achieves a threshold \cite{Criger18} that is close to the theoretically maximum possible value derived from statistical mechanics arguments \cite{Dennis02}. For random QLDPC codes with the expansion property \cite{leverrier2015quantum,fawzi2018constant,fawzi2018efficient}, the small set-flip (SSF) decoder has a theoretically proven threshold \cite{fawzi2018constant} that has been verified numerically  \cite{grospellier2018numerical}. Furthermore, in a recent study by Grospellier et al. \cite{grospellier2020combining}, it was shown that the performance of the SSF decoder can be improved by combining it with the classical BP algorithm. This two-stage BP+SSF decoder exhibits a higher code threshold, in addition to being applicable to a wider range of random QLDPC codes than the SSF decoder alone. 

In this paper, we consider another two-stage quantum decoder, first proposed by Panteleev and Kalechev \cite{Panteleev_2019}, that combines BP with a post-processing technique known as ordered statistics decoding (OSD) \cite{Fossorier_1995,fossorier2001iterative}. Panteleev and Kalechev demonstrated that for many random QLDPC codes, the BP+OSD method improves decoding performance by several orders magnitude over the BP algorithm alone. In this work, we expand on the results of Panteleev and Kalachev to provide further evidence that the BP+OSD decoder is a general decoder for all QLDPC codes that can be constructed from the hypergraph product. To this end, we first propose a new class of semi-topological codes which share properties of both topological and random QLDPC codes. We use this new class of codes to define a spectrum of QLDPC codes, and run numerical simulations to show that the BP+OSD decoder applies generally across it.

Topological QLDPC codes have stabilizers \etc{that can be locally embedded in some $D$-dimensional space} \cite{Kitaev03}. The simplest example is the surface code, obtained by taking the hypergraph product of the classical repetition code. The stabilizers of the surface code are \textit{local}, meaning they can be implemented via nearest-neighbour interactions on a 2D array of qubits \cite{Kitaev03,Fowler12}. With regards to experimental implementation, this is highly beneficial, as many qubit technologies are limited in terms of connectivity between qubits \cite{Kelly16,Rigetti16,Takita17}. Another practical advantage of the surface code is that it has a high threshold \cite{Dennis02,Fowler12,Criger18}. The disadvantage of the surface code, and topological codes in general, is that they have poor encoding rate. The surface code, for example, encodes only a single qubit per logical block meaning its encoding rate tends to zero as the code distance is increased.

Random QLDPC codes are constructed by taking the hypergraph product of high-performance classical LDPC codes. The strength of QLDPC codes over topological codes is that they can have considerably higher encoding rates that  do not tend to zero with increasing block length \cite{leverrier2015quantum,fawzi2018constant,fawzi2018efficient}. The trade-off is that random QLDPC codes have \textit{non-local} stabilizer checks, typically requiring interactions between arbitrary qubit pairs. Quantum computers based on ion traps \cite{Randall15,Debnath16,Brandl16,Ballance16}, photonic qubits \cite{Qiang18,Wang2016} or nitrogen vacancy centres \cite{wu2019programmable} promise connectivity beyond nearest-neighbours. However, such prototype devices do not yet meet the connectivity requirements of high-rate random QLDPC codes. Another disadvantage of random QLDPC codes is that they appear to have lower thresholds than their topological counterparts \cite{grospellier2018numerical,kovalev2018numerical,liu2019neural,Panteleev_2019,grospellier2020combining}. 

The new class of semi-topological codes we propose in this work allow for interpolation between local topological codes and non-local random QLDPC codes. The construction of semi-topological codes begins by modifying a classical parent code via a process called \textit{edge-augmentation}. This involves replacing each parity check edge with a length-$g$ section of repetition code referred to as a \textit{chain segment}. The semi-topological code is then obtained from the augmented parent code via the hypergraph product, which maps each of the chain-segments to a surface code-like patch. A semi-topological code can therefore be thought of as a set of surface code patches connected to one another at their boundaries via a small number of long-range interactions. The locality of a semi-topological code can be finely controlled by varying the degree to which the parent code is augmented. The ability to control connectivity makes semi-topological codes promising candidates for networked surface code architectures \cite{Nickerson14}.

In its unmodified form, the BP algorithm is ineffective for decoding QLDPC codes due to degenerate quantum errors. Quantum degeneracy is a uniquely quantum effect, and arises in situations where quantum superposition permits multiple equivalent solutions to the decoding problem. Panteleev and Kalachev \cite{Panteleev_2019} have shown that for random QLDPC codes, the problem of quantum degeneracy can be resolved by decoding using BP in conjunction with OSD post-processing. The OSD method is called when BP fails, and uses matrix inversion to resolve ambiguities in the decoding due to quantum degeneracy. 

In this work, we show that in addition to random QLDPC codes, BP+OSD enables high-performance decoding of both topological QLDPC codes and our new class of semi-topological codes. To this end, we first run numerical simulations of BP+OSD on the Toric code with increasing code distances. Our results indicate a threshold in the region $9.9\pm0.2\%$, in addition to showing evidence of exponential suppression in the low error regime. This BP+OSD threshold improves upon previous BP-based decoders for the Toric code \cite{liu2019neural}, and is close to the value of $10.3\%$ achieved by state-of-the-art decoders for the Toric code based on the minimum-weight perfect matching algorithm \cite{Edmonds65,Kolmogorov09,Criger18}. We perform further numerical simulations of BP+OSD applied to a family of semi-topological codes, as well as a family of finite-rate random QLDPC codes. For large block sizes, the BP+OSD threshold obtained for the semi-topological codes approaches the value obtained for the Toric code, reflecting the fact that the majority of stabilizer checks are 2D local.

This paper is structured as follows. In section \ref{sec:ldpc}, we first review essential concepts in classical coding theory, before introducing the edge-augmentation procedure. Section \ref{sec:quantum_coding} covers the basics of quantum stabilizer codes, and explains how they can be represented as binary linear codes. In section \ref{sec:qldpc}, we describe how QLDPC codes are obtained from classical LDPC codes via the hypergraph product, giving explicit examples of the construction of topological and random QLDPC codes. Following this, we explain how semi-topological codes are constructed by taking the hypergraph product of augmented parent codes. In section \ref{sec:bp} we describe the workings of the BP+OSD decoder. In section \ref{sec:numerics}, we describe the `combination sweep' strategy as a greedy search method for finding higher order solutions to BP+OSD. Following this, we present the results of our numerical simulations of the BP+OSD decoder for topological QLDPC codes, semi-topological codes and random QLDPC codes. Finally, in section \ref{sec:summary} we summarise and discuss directions for future work.

\section{Low density parity check codes} \label{sec:ldpc}

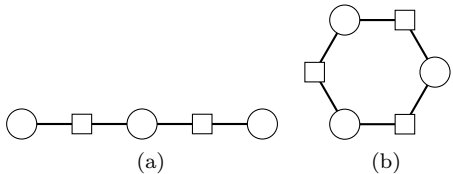
\textit{Classical error correction} --- A classical error correction code $\mathcal{C}_H$ describes a redundant encoding $\mathbf{b}\mapsto \mathbf{c}$ from a $k$-bit data string $\mathbf{b}$ to an $n$-bit codeword $\mathbf{c}$ (where $n>k$). The codewords $\mathbf{c}\in\mathcal{C}_H$ are defined as the nullspace vectors of an $m\times n$ binary parity check matrix $H$ such that $H\cdot \mathbf{c}\mod{2}=\mathbf{0}$\footnote{From this point on, we assume all arithmetic is performed modulo-2}. By the rank-nullity theorem, a parity matrix permits $k=n-\text{\sc rank}(H)$ linearly-independent codewords. If a codeword is subject to an error $\mathbf{e}$, the parity check matrix yields an $m$-bit syndrome $\mathbf{s}=H\cdot(\mathbf{c}+\mathbf{e})=H\cdot \mathbf{e}$. The syndrome will be non-zero for all errors of Hamming weight less than the code distance $|\mathbf{e}|<d$. In general, classical codes are labelled with the $[n,k,d]$ notation, where $n$ is the codeword length, $k$ is the number of encoded bits and $d$ is the code distance. The code rate is given by the ratio $R=k/n$.

\begin{figure}

	\subfloat[]{\begin{tikzpicture}
\draw[black,thick,solid] (0,0) -- (3.2,0);
\filldraw[fill=white, draw=black] (0.0,0) circle (0.2) node [label=above right:{}](D35550){};
\filldraw[fill=white, draw=black] (1.6,0) circle (0.2) node [label=above right:{}](D86454){};
\filldraw[fill=white, draw=black] (3.2,0) circle (0.2) node [label=above right:{}](D75500){};
\filldraw[fill=white, draw=black] (0.67,-0.13) rectangle (0.93,0.13) node [label={[label distance=-0.3cm]30:{}}](A84101){};
\filldraw[fill=white, draw=black] (2.2700000000000005,-0.13) rectangle (2.5300000000000002,0.13) node [label={[label distance=-0.3cm]30:{}}](A96629){};
\end{tikzpicture}}
	\subfloat[]{ \begin{tikzpicture}
\draw[black,thick,solid] (0.8,0.0) -- (0.40000000000000013,0.6928203230275509);
\draw[black,thick,solid] (0.8,0.0) -- (0.39999999999999947,-0.6928203230275513);
\draw[black,thick,solid] (-0.39999999999999986,0.692820323027551) -- (-0.8,9.797174393178826e-17);
\draw[black,thick,solid] (-0.39999999999999986,0.692820323027551) -- (0.40000000000000013,0.6928203230275509);
\draw[black,thick,solid] (-0.40000000000000036,-0.6928203230275507) -- (0.39999999999999947,-0.6928203230275513);
\draw[black,thick,solid] (-0.40000000000000036,-0.6928203230275507) -- (-0.8,9.797174393178826e-17);
\filldraw[fill=white, draw=black] (0.8,0.0) circle (0.2) node [label=above right:{}](D94057){};
\filldraw[fill=white, draw=black] (-0.39999999999999986,0.692820323027551) circle (0.2) node [label=above right:{}](D90343){};
\filldraw[fill=white, draw=black] (-0.40000000000000036,-0.6928203230275507) circle (0.2) node [label=above right:{}](D82275){};
\filldraw[fill=white, draw=black] (0.27000000000000013,0.5628203230275509) rectangle (0.5300000000000001,0.8228203230275509) node [label={[label distance=-0.3cm]30:{}}](A68680){};
\filldraw[fill=white, draw=black] (-0.93,-0.1299999999999999) rectangle (-0.67,0.13000000000000012) node [label={[label distance=-0.3cm]30:{}}](A86537){};
\filldraw[fill=white, draw=black] (0.26999999999999946,-0.8228203230275513) rectangle (0.5299999999999995,-0.5628203230275513) node [label={[label distance=-0.3cm]30:{}}](A25026){};
\end{tikzpicture} }
	
\caption{Factor graphs for two instances of the three-bit repetition code. Data nodes are drawn as circles, parity nodes as squares and edges as solid black lines. (a) The full-rank $[3,1,3]$ repetition code with parity check matrix $H=\left( \begin{smallmatrix}1&1&0\\0&1&1\end{smallmatrix} \right)$; (b) The closed-loop $[3,1,3]$ repetition code (also known as the ring code) with parity check matrix $H=\left(\begin{smallmatrix} 1&1&0\\0&1&1\\1&0&1 \end{smallmatrix}\right)$.}
\label{fig:factor_graphs}	
\end{figure}

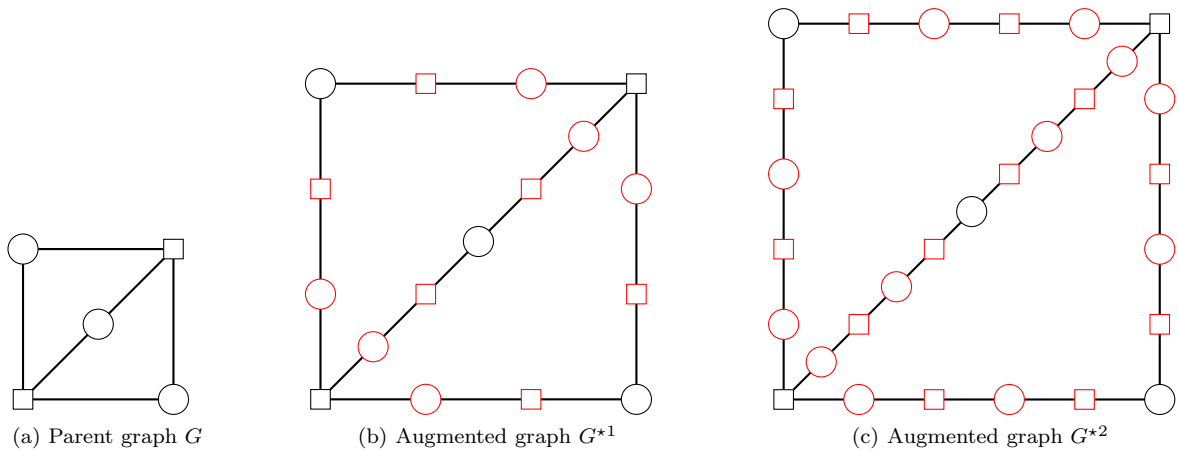
\begin{figure*} 
	
	\subfloat[Parent graph $G$]{\begin{tikzpicture}
\draw[black,thick,solid] (-1,-1) -- (1,1);
\draw[black,thick,solid] (-1,1) -- (1,1);
\draw[black,thick,solid] (1,-1) -- (1,1);
\draw[black,thick,solid] (-1,1) -- (-1,-1);
\draw[black,thick,solid] (1,-1) -- (-1,-1);
\begin{scope}[rotate=-0]\filldraw[fill=white, draw=black] (0,0) circle (0.2) node [label=above right:{}](D95211){};\end{scope}
\begin{scope}[rotate=-0]\filldraw[fill=white, draw=black] (-1,1) circle (0.2) node [label=above right:{}](D216){};\end{scope}
\begin{scope}[rotate=-0]\filldraw[fill=white, draw=black] (1,-1) circle (0.2) node [label=above right:{}](D11309){};\end{scope}
\begin{scope}[rotate=-0]\filldraw[fill=white, draw=black] (0.87,0.87) rectangle (1.13,1.13) node [label={[label distance=-0.3cm]30:{}}](A29187){};\end{scope}
\begin{scope}[rotate=-0]\filldraw[fill=white, draw=black] (-1.13,-1.13) rectangle (-0.87,-0.87) node [label={[label distance=-0.3cm]30:{}}](A98120){};\end{scope}
\end{tikzpicture}}\qquad \qquad
	\subfloat[Augmented graph $G^{\star 1}$]{\begin{tikzpicture}
\draw[black,thick,solid] (-2.0999999999999996,-2.0999999999999996) -- (2.0999999999999996,2.0999999999999996);
\draw[black,thick,solid] (-2.0999999999999996,2.0999999999999996) -- (2.0999999999999996,2.0999999999999996);
\draw[black,thick,solid] (2.0999999999999996,-2.0999999999999996) -- (2.0999999999999996,2.0999999999999996);
\draw[black,thick,solid] (-2.0999999999999996,2.0999999999999996) -- (-2.0999999999999996,-2.0999999999999996);
\draw[black,thick,solid] (2.0999999999999996,-2.0999999999999996) -- (-2.0999999999999996,-2.0999999999999996);
\begin{scope}[rotate=-0]\filldraw[fill=white, draw=black] (0,0) circle (0.2) node [label=above right:{}](D1638){};\end{scope}
\begin{scope}[rotate=-0]\filldraw[fill=white, draw=black] (-2.0999999999999996,2.0999999999999996) circle (0.2) node [label=above right:{}](D40912){};\end{scope}
\begin{scope}[rotate=-0]\filldraw[fill=white, draw=black] (2.0999999999999996,-2.0999999999999996) circle (0.2) node [label=above right:{}](D47727){};\end{scope}
\begin{scope}[rotate=-0]\filldraw[fill=white, draw=red] (1.3999999999999997,1.3999999999999997) circle (0.2) node [label=above right:{}](D26899){};\end{scope}
\begin{scope}[rotate=-0]\filldraw[fill=white, draw=red] (-1.3999999999999997,-1.3999999999999997) circle (0.2) node [label=above right:{}](D90075){};\end{scope}
\begin{scope}[rotate=-0]\filldraw[fill=white, draw=red] (2.0999999999999996,0.6999999999999997) circle (0.2) node [label=above right:{}](D88626){};\end{scope}
\begin{scope}[rotate=-0]\filldraw[fill=white, draw=red] (-2.0999999999999996,-0.6999999999999997) circle (0.2) node [label=above right:{}](D64981){};\end{scope}
\begin{scope}[rotate=-0]\filldraw[fill=white, draw=red] (0.6999999999999997,2.0999999999999996) circle (0.2) node [label=above right:{}](D9424){};\end{scope}
\begin{scope}[rotate=-0]\filldraw[fill=white, draw=red] (-0.6999999999999997,-2.0999999999999996) circle (0.2) node [label=above right:{}](D85853){};\end{scope}
\begin{scope}[rotate=-0]\filldraw[fill=white, draw=black] (1.9699999999999998,1.9699999999999998) rectangle (2.2299999999999995,2.2299999999999995) node [label={[label distance=-0.3cm]30:{}}](A72886){};\end{scope}
\begin{scope}[rotate=-0]\filldraw[fill=white, draw=black] (-2.2299999999999995,-2.2299999999999995) rectangle (-1.9699999999999998,-1.9699999999999998) node [label={[label distance=-0.3cm]30:{}}](A84120){};\end{scope}
\begin{scope}[rotate=-0]\filldraw[fill=white, draw=red] (0.5699999999999998,0.5699999999999998) rectangle (0.8299999999999998,0.8299999999999998) node [label={[label distance=-0.3cm]30:{}}](A48350){};\end{scope}
\begin{scope}[rotate=-0]\filldraw[fill=white, draw=red] (-0.8299999999999998,-0.8299999999999998) rectangle (-0.5699999999999998,-0.5699999999999998) node [label={[label distance=-0.3cm]30:{}}](A82777){};\end{scope}
\begin{scope}[rotate=-0]\filldraw[fill=white, draw=red] (1.9699999999999998,-0.83) rectangle (2.2299999999999995,-0.57) node [label={[label distance=-0.3cm]30:{}}](A90974){};\end{scope}
\begin{scope}[rotate=-0]\filldraw[fill=white, draw=red] (-2.2299999999999995,0.57) rectangle (-1.9699999999999998,0.83) node [label={[label distance=-0.3cm]30:{}}](A26336){};\end{scope}
\begin{scope}[rotate=-0]\filldraw[fill=white, draw=red] (-0.83,1.9699999999999998) rectangle (-0.57,2.2299999999999995) node [label={[label distance=-0.3cm]30:{}}](A98431){};\end{scope}
\begin{scope}[rotate=-0]\filldraw[fill=white, draw=red] (0.57,-2.2299999999999995) rectangle (0.83,-1.9699999999999998) node [label={[label distance=-0.3cm]30:{}}](A92862){};\end{scope}
\end{tikzpicture}}\qquad \qquad
	\subfloat[Augmented graph $G^{\star 2}$]{\begin{tikzpicture}
\draw[black,thick,solid] (-2.5,-2.5) -- (2.5,2.5);
\draw[black,thick,solid] (-2.5,2.5) -- (2.5,2.5);
\draw[black,thick,solid] (2.5,-2.5) -- (2.5,2.5);
\draw[black,thick,solid] (-2.5,2.5) -- (-2.5,-2.5);
\draw[black,thick,solid] (2.5,-2.5) -- (-2.5,-2.5);
\begin{scope}[rotate=-0]\filldraw[fill=white, draw=black] (0,0) circle (0.2) node [label=above right:{}](D90880){};\end{scope}
\begin{scope}[rotate=-0]\filldraw[fill=white, draw=black] (-2.5,2.5) circle (0.2) node [label=above right:{}](D11163){};\end{scope}
\begin{scope}[rotate=-0]\filldraw[fill=white, draw=black] (2.5,-2.5) circle (0.2) node [label=above right:{}](D15596){};\end{scope}
\begin{scope}[rotate=-0]\filldraw[fill=white, draw=red] (1.0,1.0) circle (0.2) node [label=above right:{}](D72579){};\end{scope}
\begin{scope}[rotate=-0]\filldraw[fill=white, draw=red] (-1.0,-1.0) circle (0.2) node [label=above right:{}](D93169){};\end{scope}
\begin{scope}[rotate=-0]\filldraw[fill=white, draw=red] (2.0,2.0) circle (0.2) node [label=above right:{}](D27146){};\end{scope}
\begin{scope}[rotate=-0]\filldraw[fill=white, draw=red] (-2.0,-2.0) circle (0.2) node [label=above right:{}](D30326){};\end{scope}
\begin{scope}[rotate=-0]\filldraw[fill=white, draw=red] (2.5,-0.5) circle (0.2) node [label=above right:{}](D19802){};\end{scope}
\begin{scope}[rotate=-0]\filldraw[fill=white, draw=red] (-2.5,0.5) circle (0.2) node [label=above right:{}](D86524){};\end{scope}
\begin{scope}[rotate=-0]\filldraw[fill=white, draw=red] (-0.5,2.5) circle (0.2) node [label=above right:{}](D82192){};\end{scope}
\begin{scope}[rotate=-0]\filldraw[fill=white, draw=red] (0.5,-2.5) circle (0.2) node [label=above right:{}](D76436){};\end{scope}
\begin{scope}[rotate=-0]\filldraw[fill=white, draw=red] (2.5,1.5) circle (0.2) node [label=above right:{}](D57017){};\end{scope}
\begin{scope}[rotate=-0]\filldraw[fill=white, draw=red] (-2.5,-1.5) circle (0.2) node [label=above right:{}](D18){};\end{scope}
\begin{scope}[rotate=-0]\filldraw[fill=white, draw=red] (1.5,2.5) circle (0.2) node [label=above right:{}](D62760){};\end{scope}
\begin{scope}[rotate=-0]\filldraw[fill=white, draw=red] (-1.5,-2.5) circle (0.2) node [label=above right:{}](D87947){};\end{scope}
\begin{scope}[rotate=-0]\filldraw[fill=white, draw=black] (2.37,2.37) rectangle (2.63,2.63) node [label={[label distance=-0.3cm]30:{}}](A94874){};\end{scope}
\begin{scope}[rotate=-0]\filldraw[fill=white, draw=black] (-2.63,-2.63) rectangle (-2.37,-2.37) node [label={[label distance=-0.3cm]30:{}}](A25357){};\end{scope}
\begin{scope}[rotate=-0]\filldraw[fill=white, draw=red] (0.37,0.37) rectangle (0.63,0.63) node [label={[label distance=-0.3cm]30:{}}](A81282){};\end{scope}
\begin{scope}[rotate=-0]\filldraw[fill=white, draw=red] (-0.63,-0.63) rectangle (-0.37,-0.37) node [label={[label distance=-0.3cm]30:{}}](A96833){};\end{scope}
\begin{scope}[rotate=-0]\filldraw[fill=white, draw=red] (1.37,1.37) rectangle (1.63,1.63) node [label={[label distance=-0.3cm]30:{}}](A42780){};\end{scope}
\begin{scope}[rotate=-0]\filldraw[fill=white, draw=red] (-1.63,-1.63) rectangle (-1.37,-1.37) node [label={[label distance=-0.3cm]30:{}}](A82009){};\end{scope}
\begin{scope}[rotate=-0]\filldraw[fill=white, draw=red] (2.37,-1.63) rectangle (2.63,-1.37) node [label={[label distance=-0.3cm]30:{}}](A78058){};\end{scope}
\begin{scope}[rotate=-0]\filldraw[fill=white, draw=red] (-2.63,1.37) rectangle (-2.37,1.63) node [label={[label distance=-0.3cm]30:{}}](A2008){};\end{scope}
\begin{scope}[rotate=-0]\filldraw[fill=white, draw=red] (-1.63,2.37) rectangle (-1.37,2.63) node [label={[label distance=-0.3cm]30:{}}](A87245){};\end{scope}
\begin{scope}[rotate=-0]\filldraw[fill=white, draw=red] (1.37,-2.63) rectangle (1.63,-2.37) node [label={[label distance=-0.3cm]30:{}}](A84576){};\end{scope}
\begin{scope}[rotate=-0]\filldraw[fill=white, draw=red] (2.37,0.37) rectangle (2.63,0.63) node [label={[label distance=-0.3cm]30:{}}](A486){};\end{scope}
\begin{scope}[rotate=-0]\filldraw[fill=white, draw=red] (-2.63,-0.63) rectangle (-2.37,-0.37) node [label={[label distance=-0.3cm]30:{}}](A74666){};\end{scope}
\begin{scope}[rotate=-0]\filldraw[fill=white, draw=red] (0.37,2.37) rectangle (0.63,2.63) node [label={[label distance=-0.3cm]30:{}}](A94276){};\end{scope}
\begin{scope}[rotate=-0]\filldraw[fill=white, draw=red] (-0.63,-2.63) rectangle (-0.37,-2.37) node [label={[label distance=-0.3cm]30:{}}](A63621){};\end{scope}
\end{tikzpicture}}

\caption{Augmented $(2,3)$-LDPC codes. (a): The parent factor graph $G$ with parity check matrix $H=\left(\begin{smallmatrix}1 & 1 &1\\1&1&1\end{smallmatrix}\right)$. (b): the $g$-augmented graph $G^{\star 1}$ with $g=1$ corresponding to a $[9,2,6]$ code; (c): the $g$-augmented code $G^{\star g}$ with $g=2$ corresponding to a $[15,2,10]$ code. The nodes belonging to the graph chain segments that form each augmented edge are coloured red.}  \label{fig:aug_codes}
\end{figure*}

\textit{Factor graphs} --- The factor graph of an $[n,k,d]$ classical code is a bipartite graph $G=(V,U,\Lambda)$ with an adjacency matrix given by the code's parity check matrix $H$ \cite{tanner1981recursive}. For an $m\times n$ parity check matrix $H$, the two sets of nodes in $G$ are defined as follows: 1) Data nodes $V=\{v_j | j=1,...,n\}$ corresponding to the columns of $H$ and taking the bit-values of the error $\mathbf{e}$; 2) Parity nodes $U=\{u_i| i=1,...,m\}$ corresponding to rows of $H$ and taking the bit-values of the syndrome $\mathbf{s}=H\cdot \mathbf{e}$. A graph edge $\lambda_{ij}\in\Lambda$ is drawn between a pair of nodes $\{v_j, u_i\}$ if $H_{ij}=1$. Factor graphs serve as a useful visualisation of the parity check matrix with applications in code design and decoding \cite{kschischang2001factor,roffe2020quantum}. Diagrammatically, factor graphs are drawn with circles representing data nodes, squares representing parity nodes and solid-lines representing the edges. Figure.~\ref{fig:factor_graphs} shows factor graphs for two instances of the three-bit repetition code.

\textit{Low density parity check (LDPC) codes} --- A family of ($l$,$q$)-LDPC codes is defined as a set of codes whose parity check matrices have column and row weights upper bounded by $l$ and $q$ respectively. As first demonstrated by Gallager \cite{gallager1962}, it is possible to construct an ($l$,$q$)-LDPC code by randomly generating a parity check matrix with the desired column and row weights.  An alternative to random LDPC code search is to employ graphical constructions in which an LDPC code family is obtained by systematically modifying the factor graph of a base code.

\textit{Edge augmented LDPC codes} --- We now introduce `edge augmentation' as a graphical method for creating an LDPC code family from the starting point of any `parent' factor graph $G=(V,U,\Lambda)$. In section \ref{sec:quantum_coding}, we show how semi-topological codes are created by taking the hypergraph product of such augmented codes.

Focusing first on a single edge $\lambda_{ij}$ connecting nodes $\{v_j,u_i\}$ in the parent code, the edge augmentation operation involves the addition of a `graph chain segment' $G^{ g}=\{V^{g},U^{g},\Lambda^g\}$ containing $g$ data nodes $V^g=\{ v^g_j|j=1,...,g \}$ and $g$ parity nodes $U^g=\{ u^g_i|i=1,...,g \}$. The adjacency matrix $H^g$ of the graph chain segment has dimensions $g\times g$. Its general form is obtained by taking a size-$g$ identity matrix and adding a `1' to the right of each of the first $g-1$ entries in the diagonal. As an example, the adjacency matrix of a graph chain segment with $g=4$ is given by
\begin{equation}
H^{g=4} = \left(  \begin{matrix}
1 & 1 & 0 & 0\\
0 & 1 & 1 & 0\\
0 & 0 & 1 & 1\\
0 & 0 & 0 & 1
\end{matrix}  \right)\rm.
\end{equation}
Following addition of the graph chain segment to the parent graph $G$, the updated factor graph $G'$ is written
\begin{equation}
G'=(V\cup V^g, U\cup U^g, \Lambda \setminus\{ \lambda_{ij}\} \cup \Lambda^g \cup \Lambda^{w}  )\rm,
\end{equation} 
where $\Lambda\setminus \{\lambda_{ij} \}$ is the original parent edge set minus the edge that has been augmented. Two additional edges $\Lambda^w=\{\lambda^g_{1j}, \lambda^g_{ig}\} $ are added to connect the nodes $\{v_j,u^g_1\}$ and $\{v^g_g,u_i\}$. These edges `weld' the graph chain segment to the parent nodes $\{v_j,u_i\}$. 

A $g$-augmented factor graph $G^{\star g}=(V^{\star g},U^{\star g},\Lambda^{\star g})$ is obtained by edge-augmenting each edge in a parent graph $G=(V, U,\Lambda)$ with a length-$g$ graph chain segment. If $G$ corresponds to an $[n,k,d]$ code with parity check matrix $H$, then the $g$-augmented graph $G^{\star g}$ corresponds to an $[n+g|\Lambda|,k,d']$ code with parity check matrix $H^{\star g}$, where $|\Lambda|$ is the number of edges in the parent graph. The augmented code distance depends upon the structure of $H$, but is lower-bounded by $d'\geq (1+g\mu)d$, where $\mu$ is the minimum degree over all data nodes $V$ (for the proof of this lower bound see Appendix.~\ref{app:lower_bound}). If the parent graph $G$ is an $(l,q)$-LDPC code with $l,q\geq2$, then the g-augmented graph $G^{\star g}$ will also be an $(l,q)$-LDPC code. A family of LDPC codes with increasing code distance can be obtained by augmenting a parent code with increasing values of the augmentation parameter $g$. The tradeoff in the edge augmentation procedure is a reduction in the code rate: if the parent graph has rate $R=k/n$, the augmented graph will have rate $R^{\star g}=\frac{R}{1+g|\Lambda|/n}$. Any increases in code distance due to edge augmentation must therefore be balanced against the respective increase in code overhead.

 Figure.~\ref{fig:aug_codes} illustrates the first three levels of a ($2,3$)-LDPC code family starting from a $[3,2,2]$ parent code with parity check matrix $H=\left(\begin{smallmatrix}1 & 1 &1\\1&1&1\end{smallmatrix}\right)$. The factor graph of the parent code $G$ is shown in Figure.~\ref{fig:aug_codes}a. Figure.~\ref{fig:aug_codes}b shows the $g$-augmented graph $G^{\star 1}$ with $g=1$ and code parameters $[9,2,6]$. Here, each edge in the parent graph $G$ has been augmented with a length-$1$ graph chain segment, the nodes of which are coloured red. Figure.~\ref{fig:aug_codes}c is the $g$-augmented graph $G^{\star 2}$ corresponding to a code with parameters $[15,2,10]$.

\section{Quantum coding}\label{sec:quantum_coding}

\textit{Quantum error correction} --- Quantum bits (qubits) are susceptible to a continuum of errors corresponding to rotations about the Bloch sphere. Fortunately, due to an effect known as the digitization of the error, quantum errors can be modelled in terms of the random occurrence of a discrete set of Pauli-operators $\{\openone,X,Y,Z\}$\footnote{The Pauli operators are defined as follows: $\openone=\left(\begin{smallmatrix}1&0\\0&1\end{smallmatrix}\right)$; $X=\left(\begin{smallmatrix}0&1\\1&0\end{smallmatrix}\right)$; $Y=\left(\begin{smallmatrix}0&-\text{i}\\ \text{i}&0\end{smallmatrix}\right)$; $Z=\left(\begin{smallmatrix}1&0\\0&-1\end{smallmatrix}\right)$.} on each qubit \cite{Knill97}. An $[[n,k,d]]$ quantum error correction code $\mathcal{Q}$ is a mapping $\ket{\psi}\mapsto \ket{\psi}_L$ from a $k$-qubit quantum state $\ket{\psi}$ to an entangled $n$-qubit codeword (logical) state $\ket{\psi}_L$. The quantum codewords $\ket{\psi}_L\in \mathcal{Q}$ satisfy the condition $S_j\ket{\psi}_L=(+1)\ket{\psi}_L$ for all $S_j\in\mathcal{S}$, where $\mathcal{S}$ is a group of mutually commuting Pauli operators known as the code's stabilizer \cite{Gottesman97}. Pauli-errors of Hamming weight less than the code distance $|E|<d$ will result in at least one stabilizer $S_k$ projecting onto the negative eigenspace $S_k\ket{\psi}_L=(-1)\ket{\psi}_L$. 

The Pauli group has a convenient binary representation in which each operator is mapped to a length-$2$ vector: $\openone\mapsto (0 , 0)$, $X\mapsto (1,0)$, $Z\mapsto(0,1)$ and $Y\mapsto(1,1)$. In general, the binary representation of an $n$-qubit Pauli operator $K$ will be a length-$2n$ vector of the form $\mathbf{k} = (\mathbf{x},\mathbf{z})$, where $\mathbf{x}$ and $\mathbf{z}$ both have length $n$ and represent the positions of $X$- and $Z$-Pauli components respectively. As an example, the binary representation of the three-qubit Pauli operator $K=X_1Z_3$ is $\mathbf{k}=(100,001)$. The binary representation provides a useful setting from which to repurpose existing classical coding techniques for quantum error correction.

A quantum parity check matrix is defined as a matrix in which each row corresponds to a code stabilizer in its binary representation. Calderbank, Shor and Steane (CSS) codes \cite{Calderbank95,Steane96b,Steane97} are a subset of quantum codes with parity check matrices of the form $H_{\rm CSS}=\left(\begin{smallmatrix}H_Z & 0\\0&H_X\end{smallmatrix}\right)$, where $H_Z\cdot H_X^T=\mathbf{0}$ due to the requirement that the stabilizers commute. For a CSS code subject to a Pauli error $E\mapsto \mathbf{e}_Q=(\mathbf{x},\mathbf{z})$, the quantum syndrome $\mathbf{s}_Q$ is calculated as follows 
\begin{equation}
\mathbf{s}_Q=(\mathbf{s}_X,\mathbf{s}_Z)=(H_Z\cdot\mathbf{x}, \ H_X\cdot \mathbf{z})\rm.
\end{equation} From the above, it can be seen that the working of a CSS code can be thought of in terms of two classical codes, $\mathcal{C}(H_Z)$ and $\mathcal{C}(H_X)$, designed to detect bit-flips ($X$-errors) and phase-flips ($Z$-errors) respectively. 

\textit{Hypergraph product codes} -- The hypergraph product, first proposed by Tillich and Zemor \cite{tillich2013quantum}, is a method for converting classical code pairs $\{\mathcal{C}_{H_1},\mathcal{C}_{H_2}\}$ to a quantum CSS code $\mathcal{HGP}(\mathcal{C}_{H_1},\mathcal{C}_{H_2})$. In the below, we describe the special case of the symmetric hypergraph product $\mathcal{HGP}(\mathcal{C}_H)$ for which $\mathcal{C}_{H_2}=\mathcal{C}_{H_1}$. 

For a classical code $\mathcal{C}_H$ with code parameters $[n,k,d]$, the symmetric product $\mathcal{HGP}(\mathcal{C}_H)$ is a CSS code with
\begin{equation}\label{eq:hgp_h}
\begin{split}
H_X=( \ H \otimes \openone_{n} \ | \ \openone_{m} \otimes H^T \ ){\rm,}\\
H_Z = (\ \openone_{n} \otimes H \ | \ H^T \otimes \openone_{m} \ )\rm,
\end{split}
\end{equation}
where $H^T$ is the transpose parity check matrix describing a `transpose' code $\mathcal{C}_{H}^T$ with parameters $[m,k^T,d^T]$. Here, $k^T$ is the number of logical qubits encoded by the transpose code whilst $d^T$ is the distance of the transpose code.  The quantum code parameters of $\mathcal{HGP}(H)$ are \begin{equation}\label{eq:hgp_params}
[[n^2+m^2,\ k^2+(k^T)^2,\ {\text{\sc min}}(d,d^T)]]\rm.
\end{equation}
The specific advantage of the hypergraph product construction is that it allows \textit{any} classical code to be converted to a quantum code: the commutativity constraint $H_Z\cdot H_X^T=\mathbf{0}$ is satisfied for all  binary parity check matrices $H$.

\section{Quantum LDPC (QLDPC) codes}\label{sec:qldpc}

An  $(l_Q,q_Q)$-QLDPC code family is defined as a set of CSS codes whose quantum parity check matrices $H_{\rm CSS}$ have row and column weights upper bounded by $l_Q$ and $q_Q$ respectively \cite{mackay2004sparse}. \etc{The hypergraph product preserves the sparsity of the original classical code} \cite{tillich2013quantum}.  From the structure of equation (\ref{eq:hgp_h}), we see that the hypergraph product of an ($l,q$)-LDPC code with parity check matrix $H$ results in an ($l_Q$,$q_Q$)-QLDPC code with quantum parity check matrix $H_Q$, where $l_Q=\text{\sc max}(2l,2q)$ and $q_Q=l+q$.  The hypergraph product of a classical LDPC code family is therefore a quantum LDPC (QLDPC) code family.

\etc{Two important classes of hypergraph product codes are:} 1) topological QLDPC codes, such as the surface and Toric codes, constructed by taking the hypergraph product of repetition codes; 2) random QLDPC codes constructed by taking the hypergraph product of randomly generated classical LDPC codes. When random codes generate a factor graph with the expansion property, these are known as `quantum expander codes' \cite{leverrier2015quantum,fawzi2018constant,fawzi2018efficient}. In this section, we propose a new class of semi-topological codes constructed by taking the hypergraph product of augmented LDPC code families. Semi-topological codes are designed to share properties of both random and topological QLDPC codes.

\textit{Topological $(4,4)$-QLDPC codes} --- The hypergraph product of an $[n,1,n]$ full-rank repetition code (see Figure.~\ref{fig:factor_graphs}a for an example) yields a surface code with parameters $[[n^2+(n-1)^2,1,n]]$. Likewise, the hypergraph product of the closed-loop repetition code (also known as the ring code, see Figure.~\ref{fig:factor_graphs}b for an example) results in a Toric code with parameters $[[2n^2,2,n]]$. Topological codes such as the surface code are considered leading candidates for experiment due to their high threshold \cite{Dennis02} and the fact that they are local: all code stabilizers can be measured via interactions between nearest neighbour qubits \cite{Fowler12}. Another advantage of the topological codes is that they have parity check matrices that are $(4,4)$-QLDPC, meaning each stabilizer measurement involves at most four qubits. From a hardware perspective, this is beneficial, \etc{as each parity check operation} involves error-prone multi-qubit operations. The shortcoming of topological codes is that they scale poorly in terms of rate: $R=k/n \rightarrow 0$ as $d$ is increased.

\begin{table} 
\renewcommand{\arraystretch}{1.2}

\begin{tabular}{c c | c c c}
	$\mathcal{C}_H$ & $\mathcal{C}_H^T$& $\mathcal{HGP}(\mathcal{C}_H)$ & $R=k/n$ & $\bar{w}$ \\ \hline
	$[16,4,6]$  & $[12,0,\infty]$ & $[[400,16,6]]$ & $0.04$ & $7.0$\\
	$[20,5,8]$  & $[15,0,\infty]$ & $[[625,25,8]]$ & $ 0.04$ & $7.0$\\
	$[24,6,10]$ & $[18,0,\infty]$ & $[[900,36,10]]$ & $0.04$ & $7.0$\\
\end{tabular}

\caption{A constant-rate $(8,7)$-QLDPC code family $\mathcal{HGP}(\mathcal{C}_H)$ constructed from the hypergraph product of classical $(3,4)$-LDPC codes $\mathcal{C}_H$. Column 1: $[n,k,d]$ parameters of the classical $(3,4)$-LDPC codes $\mathcal{C}_H$. The parity check matrices of these codes have full rank. Column 2: $[m,k^T,d^T]$ parameters of the transpose codes $\mathcal{C}_H^T$. The distance of these codes is set to $d^T=\infty$ as they encode zero logical bits. Column 3: $[[n,k,d]]$ parameters of the $(8,7)$-QLDPC codes $\mathcal{HGP}(\mathcal{C}_H)$. Column 4: The rate of the QLDPC code $\mathcal{HGP}(\mathcal{C}_H)$. Column 5: the average check weight $\bar{w}$ of $\mathcal{HGP}(\mathcal{C}_H)$.}
\label{tab:qldpc}

\end{table}

\textit{Random QLDPC codes} --- Random QLDPC codes are constructed from the hypergraph product of randomly generated classical LDPC codes \cite{kovalev2018numerical}. The advantage of random QLDPC codes, over topological codes, is that they can encode more qubits per logical block. Table.~\ref{tab:qldpc} lists members of an $(8,7)$-QLDPC code family constructed by taking the hypergraph product of a family of randomly generated $(3,4)$-LDPC codes. The $(3,4)$-LDPC classical code family was obtained using the Mackay-Neal method which ensures the randomly generated parity check matrix has no length-four cycles \cite{mackay1997near}. The resultant $(8,7)$-QLDPC hypergraph product codes are finite-rate, with $R=k/n=0.04$ as the distance is increased. The disadvantage of QLDPC codes is that they are highly non-local, requiring arbitrary qubit-qubit interconnectivity to perform stabilizer checks. Furthermore, the stabilizers typically involve more qubits than topological codes. The family of codes shown in Table.~\ref{tab:qldpc}, for example, are $(8,7)$-QLDPC with stabilizer checks of mean weight $\bar{w}=7.0$. This is higher than the mean check weight of $\bar{w}=4.0$ for the $(4,4)$-QLDPC Toric codes.

\textit{Semi-topological codes} -- Semi-topological codes are constructed by taking the hypergraph product of augmented LDPC codes. Table.~\ref{tab:aug_codes} shows the code parameters of a family of semi-topological codes constructed from $(2,3)$-LDPC augmented codes of the type illustrated in Figure.~\ref{fig:aug_codes}. For an augmented code $\mathcal{C}_H^{*g}$, each augmented edge can be thought of as a section of a repetition code. The hypergraph product $\mathcal{HGP}(\mathcal{C}_H^{*g})$ therefore maps each augmented edge to a section of code that resembles a surface code. In these regions, the code stabilizers will be local. As the distance of the augmented code is increased, the resultant semi-topological code contains larger surface code patches and becomes more local in nature. This convergence to surface code-like structure is shown by the check-weight parameter $\bar{w}$ in Table.~\ref{tab:aug_codes}, which tends to $4.0$ with increasing code distance as the local surface code-like patches begin to dominate. We term this new family `semi-topological codes', as they encode more logical qubits than the topological codes whilst requiring fewer long range interactions than random QLDPC codes.

\begin{table} 
	\renewcommand{\arraystretch}{1.2}
	
	\begin{tabular}{c c c | c c c}
		$g$ & $\mathcal{C}_H^{*g}$ & $(\mathcal{C}_H^{*g})^T$ & $\mathcal{HGP}(\mathcal{C}_H^{*g})$ & $R$ & $\bar{w}$ \\ \hline
		0  &$[3,2,2]$ & $[2,1,1]$ & $[[13,5,2]]$ & $0.385$ & $5.00$ \\
		1 &$[9,2,6]$  & $[8,1,8]$ & $[[145,5,6]]$ & $ 0.0345$& $4.25$ \\
		2 &$[15,2,10]$ & $[14,1,14]$ & $[[421,5,10]]$ & $0.0119$& $4.14$ \\
		3 &$[21,2,14]$ & $[20,1,20]$ & $[[841,5,14]]$ & $0.00595$& $4.10$ \\
		9 &$[57,2,38]$ & $[56,1,56]$& $[[6385,5,38]]$ & $0.000783$& $4.04$ \\
	\end{tabular}
\caption{A semi-topological code family $\mathcal{HGP}(\mathcal{C}_H^{*g})$ constructed from the augmented $(2,3)$-LDPC codes $\mathcal{C}_H^{*g}$. Column 1: the code augmentation parameter $g$. Column 2: $[n,k,d]$ parameters for the augmented $(2,3)$-LDPC codes. Column 3: $[m,k^T,d^T]$ parameters of the transpose code $(\mathcal{C}_H^{*g})^T$. Column 4: $[[n,k,d]]$ parameters of the semi-topological code $\mathcal{HGP}(\mathcal{C}_H^{*g})$. These codes are $(6,5)$-QLDPC. Column 5: code rate $R=k/n$ of $\mathcal{HGP}(\mathcal{C}_H^{*g})$. Column 6: average check-weight $\bar{w}$ of $\mathcal{HGP}(\mathcal{C}_H^{*g})$.}

\label{tab:aug_codes}

\end{table}

\section{Belief propagation decoding}\label{sec:bp}

In the classical setting, the role of the decoder is to determine the most likely error-string $\mathbf{e}$ satisfying the syndrome equation $H\cdot\mathbf{e}=\mathbf{s}$. In practice, this decoding problem amounts to finding a minimum weight (MW) estimate of the error $\mathbf{e}_{\rm MW} \mapsto \text{\sc{argmax}}_\mathbf{e} \ P(\mathbf{e}|\mathbf{s})$. For a uniformly distributed random noise model, the MW estimate can be computed bit-wise by calculating the marginal probability that bit $e_i=1$ as follows
\begin{equation} \label{eq:marginals}
P_1(e_i) = \sum\nolimits_{\sim e_i } P(e_1,e_2,e_i=1,e_3,...,e_n|\mathbf{s}) 
\end{equation} 
where $\sum\nolimits_{\sim e_i }$ denotes a summation over all bits $e_j$ except $e_i$. The marginal $P_1(e_i)$ is referred to as a \textit{soft-decision} for the bit $e_i$. The final decoding estimate (\textit{hard-decision}) is then made for each bit according to
\begin{equation}
(e_{\rm MW})_i=\begin{cases}
1 & {\rm if} \ P_1(e_i)\geq 0.5 \\
0 & {\rm if} \ P_1(e_i) < 0.5
\end{cases}\rm.
\end{equation} 
Belief propagation (BP) is an efficient marginalisation algorithm and the backbone of many high-performance classical decoders \cite{mackay1999good}. The essential intuition underpinning BP is that (for certain codes) the probability distribution $P(\mathbf{e}|\mathbf{s})$ can be factorised in a way that reduces the number of repeat summations in the computation of the marginals. The specific form of this factorisation is deduced from the structure of the code's factor graph. The BP algorithm computes exact marginals when applied to codes with tree-like factor graphs. For factor graphs with loops, the BP decoder outputs approximate marginals. However, it has been shown \cite{mackay1997near} that good decoding performance is nonetheless possible provided the factor graph is sufficiently loop-free.

The BP decoder takes a parity check matrix $H$ and a syndrome $\mathbf{s}$ as input. The algorithm iteratively updates a soft-decision vector $P_1(\mathbf{e})$ by passing sets of `beliefs' between the nodes of the factor graph. At each iteration, a BP estimate $\mathbf{e}_{\rm BP}$ is obtained via a hard decision on $P_1(\mathbf{e})$. If the BP estimate satisfies the syndrome equation, $H\cdot \mathbf{e}_{\rm BP}=\mathbf{s}$, the BP decoder is said to have `converged' and the BP algorithm is terminated. The BP decoder fails if convergence does not occur within a number of iterations equal to the block length of the code. A more detailed description of BP can be found in Appendix.~\ref{app:bp}.

\textit{BP decoding of quantum codes} -- For a CSS code subject to a Pauli error $E\mapsto \mathbf{e}_Q=(\mathbf{x},\mathbf{z})$, the quantum syndrome is given by $\mathbf{s}_Q=(\mathbf{s}_x,\mathbf{s}_z)=(H_Z\cdot\mathbf{x} , \ H_X\cdot \mathbf{z})$. Assuming a Pauli-noise model with uncorrelated $X$- and $Z$-errors, the CSS code can be decoded independently as two classical codes with syndrome equations $\mathbf{s}_x=H_Z\cdot \mathbf{x}$ and $\mathbf{s}_z=H_X\cdot \mathbf{z}$. Unfortunately, the unmodified BP algorithm cannot be used to directly decode CSS codes owing (in part) to an effect known as quantum degeneracy. Quantum degeneracy arises due to the fact that there can be multiple minimum-weight solutions to the quantum decoding problem. In classical coding the goal is to estimate the exact error configuration that occurred $\mathbf{e}_{MW}=\mathbf{e}$. In contrast, for quantum coding, it is sufficient to find any recovery operation $\mathbf{r}_Q$ that is equivalent to the error up to a stabiliser $\mathbf{r}_Q + \mathbf{e}_Q = \text{\sc rowspace}(H_{CSS})$. For BP decoding, quantum degeneracy becomes problematic when are there multiple minimum-weight solutions satisfying the syndrome equation. As an example, consider a bit-error decoding problem $\mathbf{s}_x=H_Z\cdot \mathbf{x}$ that has two minimum-weight solutions $\mathbf{x}_1$ and $\mathbf{x}_2$. As the degenerate solutions have equal Hamming weight $|\mathbf{x}_1|=|\mathbf{x}_2|$ the BP decoder assigns high probability to both. This situation is referred to as a split-belief \cite{Criger18}, and leads to a BP output of the form $\mathbf{x}_{BP}=\mathbf{x}_1+\mathbf{x}_2$. In this case, $H_Z\cdot \mathbf{x}_{BP} = \mathbf{s}_x + \mathbf{s}_x= \mathbf{0} \neq \mathbf{s}_x$. The BP decoder therefore fails to converge when there are split beliefs of this type.

\textit{Ordered statistics decoding} -- Many attempts have been made to modify or supplement the BP algorithm to solve the problem of quantum degeneracy. The most successful approach to date involves applying a post-processing algorithm known as the ordered statistics decoder (OSD). Originally designed as a method for reducing error floors in classical LDPC codes by Fossosier and Lin \cite{Fossorier_1995}, OSD was first applied in the quantum setting by Panteleev and Kalachev \cite{Panteleev_2019} and shown to be a surprisingly effective decoder of random QLDPC codes.  In this paper, we show that OSD also performs well for the Toric codes and our new class of semi-topological codes. We also provide the first open-source demonstration of the algorithm \cite{bp_osd}. Note that in the below, for notational simplicity, we describe OSD post-processing as applied to a classical decoding problem $\mathbf{s}=H \cdot \mathbf{e}$. The procedure we outline applies equally to decoding the $H_X$ and $H_Z$ components of a CSS code.

As parity check matrices do not have full column-rank, it is not possible to solve the syndrome equation by matrix inversion $H^{-1}\cdot \mathbf{s}= \mathbf{e}$. However, for any parity check matrix it is possible to find a subset of columns, specified by the indices $[S]$, that are linearly independent. These columns form a basis and can be used to define a sub-matrix $H_{[S]}$ with full column-rank, formed by selecting the columns $[S]$ of the original parity check matrix $H$. As this sub-matrix has full column-rank, it can be inverted to give a solution to the syndrome equation $H_{[S]}^{-1}\cdot \mathbf{s} = \mathbf{e}_{[S]}$. Each choice of the basis $[S]$ corresponds to a unique solution $\mathbf{e}_{[S]}$, eliminating any potential ambiguity due to quantum degeneracy. It is possible to select $[S]$ as a random basis set, but this approach is unlikely to result in a good (low-weight) solution for $\mathbf{e}_{[S]}$. The idea behind the OSD post-processing algorithm is that the soft-decisions from BP are used to select a basis-set $[S]$ containing bits that have high-probability of having been flipped.

\textit{The OSD-0 algorithm} --- In a BP+OSD decoder, the OSD post-processing step is called when the BP algorithm fails to converge within a number of iterations equal to the block length of the code. The simplest manifestation of the OSD decoder is known as OSD-0, the steps of which are as follows:
\begin{enumerate}
	\item Use the BP soft decision vector $P_1(\mathbf{e})$ to obtain a ranked list of bit-indices $[O_{BP}]$ ordered (left-to-right) from most-to-least likely of being flipped.
	\item Order the columns of the parity check matrix $H_{[O_{BP}]}$ according to the ranking $[O_{BP}]$.
	\item Select the first $\text{\sc rank}{(H)}$ linearly independent columns of $H_{[O_{BP}]}$ as the most-probable basis-set $[S]$.
	\item Calculate the OSD-0 solution on the basis-bits by matrix inversion $\mathbf{e}_{[S]}=H_{[S]}^{-1}\cdot \mathbf{s}$.
	\item The OSD-0 solution across all bits is given by $\mathbf{e}_{[S,T]} = \left(\mathbf{e}_{[S]},\mathbf{e}_{[T]}\right)=\left(\mathbf{e}_{[S]},\mathbf{0}\right)$, where we define the remainder-set $[T]$ as the bits which are not in the basis-set $[T]\notin[S]$. The OSD-0 solution will always satisfy the syndrome equation $H_{[S,T]}\cdot \mathbf{e}_{[S,T]} = \mathbf{s}$.
	\item Map the OSD-0 solution to the original bit ordering $\mathbf{e}_{[S,T]}\mapsto \mathbf{e}_{\text{OSD-0}}$.
	
\end{enumerate}

\textit{Higher order OSD} --- In higher-order OSD, we consider solutions for which $\mathbf{e}_{[T]}\neq \mathbf{0}$. The first step involves computing the OSD-0 solution $\mathbf{e}_{[S]}$ on the basis bits as described above. Following this, for a given choice of $\mathbf{e}_{[T]}$, the higher order OSD solution across all bits is given by
\begin{equation}\label{eq:osd_enc}
\mathbf{e}_{[S,T]} = \left(H_{[S]}^{-1}\cdot \mathbf{e}_{[S]}+H_{[S]}^{-1}\cdot H_{[T]} \cdot \mathbf{e}_{[T]}, \ \mathbf{e}_{[T]} \right) .\end{equation} 
Note that the above solution satisfies the syndrome relation $H_{[S,T]}\cdot \mathbf{e}_{[S,T]}=\mathbf{s}$ for all possible configurations of $\mathbf{e}_{[T]}$. A higher order OSD routine involves searching over different values of $\mathbf{e}_{[T]}$ to find the OSD solution with the lowest Hamming weight $\text{\sc{min}}(|\mathbf{e}_{[S,T]}|)$. The length of the $\mathbf{e}_{[T]}$ vector is equal to $k'=n-\text{\sc rank}(H)$, meaning there are $2^{k'}$ distinct configurations: as a result, searching over all configurations soon becomes intractable for large codes. However, the BP soft-decision vector $P_1(\mathbf{e})$ can be used to rank the bits in $\mathbf{e}_{[T]}$. Good solutions can then be discovered by implementing a weighted greedy search routine which prioritises the more probable configurations of $\mathbf{e}_{[T]}$ according to the soft-decisions $P_1(\mathbf{e})$.

\textit{Greedy search strategies for higher order OSD} -- For the numerical simulations in this paper, we implement a greedy search method we refer to as the `combination sweep strategy', a variant of the method originally proposed in \cite{Fossorier_1995}. The steps of the combination sweep strategy are as follows: \begin{enumerate}
	\item The bits in the $\mathbf{e}_{[T]}$ component of the OSD solution are sorted according to the BP soft decisions.
	\item All weight-one configurations of $\mathbf{e}_{[T]}$ are searched over.
	\item All weight-two configurations in the first $\lambda$ bits of $\mathbf{e}_{[T]}$ are searched over. The total number of configurations considered is equal to $k'+{\lambda \choose 2}$, where $k'=n-\text{\sc rank}(H)$ is the length of the $\mathbf{e}_{[T]}$ vector.
\end{enumerate}
We label our decoders using the combination sweep greedy search algorithm as BP+OSD-CS. For all the simulations in this work, we set the combination sweep search depth parameter to $\lambda=60$.  Note that in \cite{Panteleev_2019}, Panteleev and Kalachev used a different greedy search method that involved testing all $2^{\lambda}$ permutations of the first $\lambda$ bits in $\textbf{x}_{[T]}$. For a fixed number of search terms, the combination sweep search algorithm provides a modest improvement in decoding performance over this exhaustive approach. For more details, see Appendix.~\ref{app:comparison}.

\section{Numerical Simulations}\label{sec:numerics}

\begin{figure}
	\input{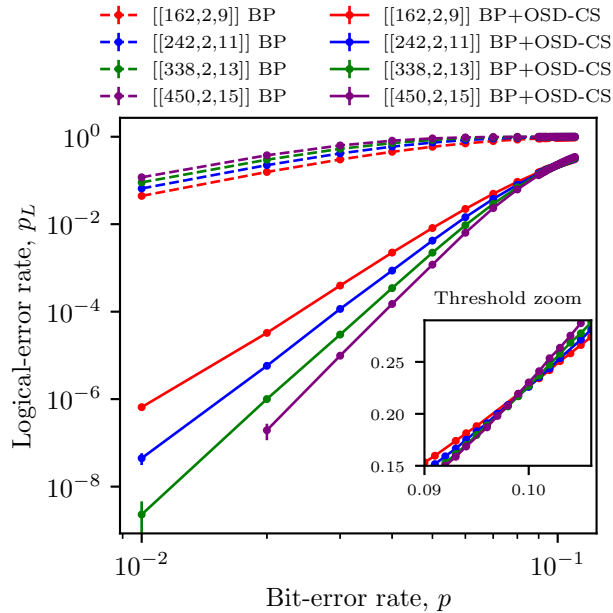}
	\caption{Toric code threshold plot comparing the BP decoder (dashed lines) versus the BP+OSD-CS decoder (solid lines). The logical error rate $p_L$ is plotted against the physical error rate $p$ for code distances $d=\{9,11,13,15\}$. For this simulation, the search depth parameter for the greedy search `combination sweep strategy' is set to $\lambda=60$.} 
	\label{plot:toric}
	
\end{figure}

\begin{figure}
	\input{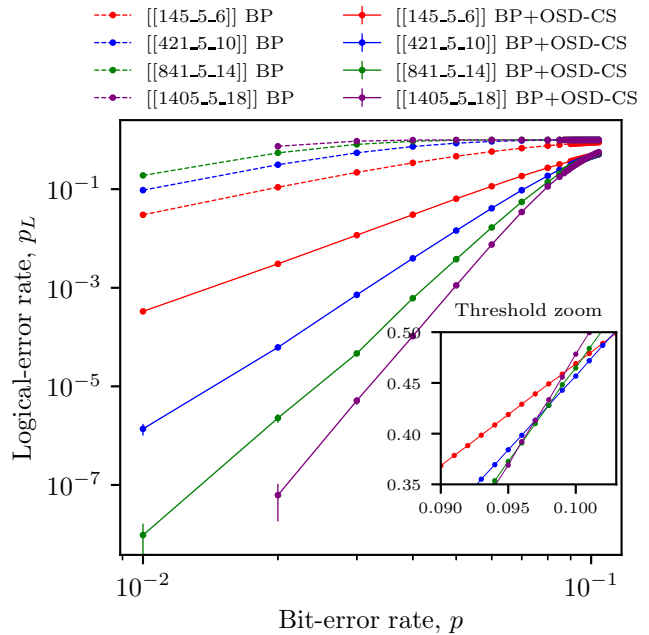}
	\caption{Threshold plot for the semi-topological codes constructed from a family of augmented codes (see Table.~\ref{tab:aug_codes} for the code parameters). The logical error rate $p_L$ is plotted against the physical error rate $p$ for code distances $d=\{6,10,14,18\}$. The search depth parameter for the greedy search combination sweep strategy is set to $\lambda=60$.} 
	\label{plot:ihgp}
	
\end{figure}

\textit{Simulation methodology for BP+OSD decoding} --- For the numerical simulations of the BP+OSD decoder in this work, we sample errors from the code capacity channel under the assumption that $X$- and $Z$-type errors are uncorrelated. As the quantum error correction codes we consider are constructed from a symmetric hypergraph product, the respective decoding problems for $X$- and $Z$-type errors are equivalent. As such, it suffices to simulate a single error species to assess decoding performance. Here, we sample $X$-errors and solve the decoding problem $\mathbf{s}_x=H_Z \cdot \mathbf{x}$. The pseudocode for the specific implementation of BP we use for the numerical simulations in this paper can be found in Appendix.~\ref{app:bp}. The simulation chain we implement for each BP+OSD decoding cycle is described below:
\begin{enumerate}
	\item An error $\mathbf{x}$ is randomly sampled from a binary symmetric channel with bit error rate $p$. The syndrome is then calculated $\mathbf{s}_x=H_Z\cdot \mathbf{x}$.
	
	\item The BP decoder is called with $H_Z$ and $\mathbf{s}$ as inputs. The output of the BP decoder is a candidate solution $\mathbf{x}_{BP}$ along with its respective soft-decision vector $P_1(\mathbf{x})$. If $H_Z\cdot \mathbf{x}_{BP} = \mathbf{s}_x$, then the BP decoder has converged and the simulation jumps directly to step $5$. If $H_Z\cdot \mathbf{x}_{BP} \neq \mathbf{s}_x$, then the OSD post-processing routine (steps 3-4) is called. For our decoding simulations we use the `min-sum' variant of BP algorithm as described in \cite{emran2014simplified}.
	
	\item The OSD-0 post-processing method, as described above, is used to obtain a solution of the form $\mathbf{x}_{[S,T]} = \left(\mathbf{x}_{[S]},\mathbf{x}_{[T]}\right)=\left(\mathbf{x}_{[S]},\mathbf{0}\right)$.
	
	\item A greedy algorithm is run to search for higher-order OSD solutions that improve upon OSD-0. For this work, we adopt the combination sweep strategy with the search depth parameter set to $\lambda=60$. However, in general, the specific form of the greedy search routine can be tailored according to parameters such as the physical error rate or code structure. The lowest weight OSD solution, $\text{\sc{min}}|\mathbf{e}_{[S,T]}|$, is mapped to the original bit-ordering and chosen as the BP+OSD candidate solution $\mathbf{e}_{\text{OSD}}$.
	
	\item After applying the recovery provided by the decoder, the `residual' error is given by $\mathbf{x}_R=\mathbf{x} + \mathbf{x}_{\rm OSD}$ (or in the case where BP converged $\mathbf{x}_R=\mathbf{x} + \mathbf{x}_{\rm BP}$). The decoding cycle is counted as a success if $\mathbf{x}_R$ is a not an $X$-type logical operator of the code. By definition, an $X$-type logical operator will anti-commute with its corresponding $Z$-type logical operator. Checking for decoding success therefore involves verifying that $L_Z\cdot \mathbf{x}_R=\mathbf{0}$, where $L_Z$ is a matrix in which each row represents a $Z$-type logical operator.

\end{enumerate}
\etc{Next, we discuss our thresholds estimates for thre code families across the QLDPC code spectrum, with an overview presented in Table.~\ref{TableThresholds}.}

\textit{Topological QLDPC codes} --- Figure.~\ref{plot:toric} shows a Toric code threshold plot comparing the BP decoder against the BP+OSD-CS decoder. The logical error rate $p_L$ is plotted against the physical error rate $p$ for code distances $d=\{9,11,13,15\}$. Due to quantum degeneracy, the BP decoder alone (dashed lines) does not exhibit a threshold: increasing the code distance $d$ increases the logical error rate $p_L$ for all values of the bit-error rate $p$. In contrast, the BP+OSD-CS decoder (solid lines) shows crossings that indicate a threshold in the region $9.9\pm0.2\%$. Furthermore, by inspection of the sub-threshold regime, we see evidence of exponential suppression in the logical error rate with decreasing physical error rate. The corresponding threshold (not plotted) for the BP+OSD-0 decoder is $9.2\pm0.2\%$. Performing the combination-sweep for higher-order OSD solutions therefore results in a quantifiable improvement in decoding performance.

\begin{figure}
	\input{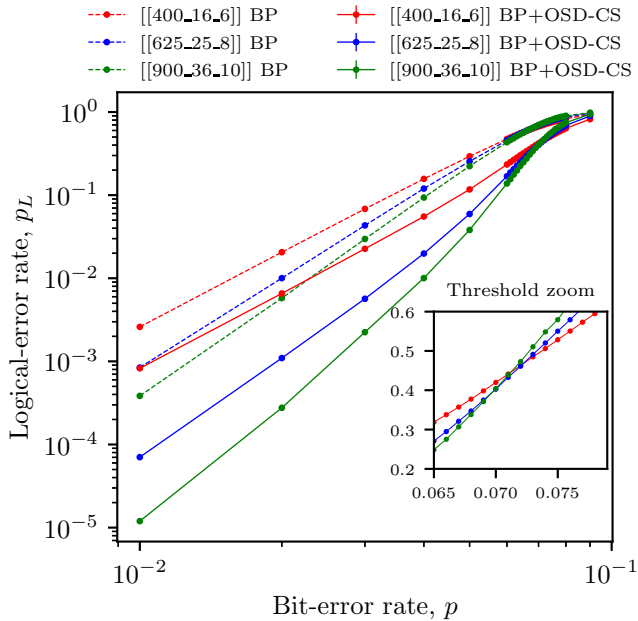}
	\caption{Threshold plots for the family of constant rate QLDPC codes listed in Table.~\ref{tab:qldpc}. The logical error rate $p_L$ is plotted against the physical error rate $p$ for code distances $d=\{6,8,10\}$ The search depth parameter for the greedy search combination sweep strategy is set to $\lambda=60$.}
	\label{plot:qldpc}
\end{figure}

\textit{Semi-topological codes} --- Figure.~\ref{plot:ihgp} shows the threshold plot for a family of semi-topological codes constructed from augmented $(2,3)$-LDPC codes. The parameters for this code family are listed in Table.~\ref{tab:aug_codes}. The logical error rates $p_L$ (for both BP and BP+OSD-CS) are plotted against the physical error rate $p$ for code distances $d=\{6,10,14,18\}$. As with the Toric codes, the BP decoder alone does not yield a threshold. For the BP+OSD-CS decoder, however, a crossing is clearly visible, suggesting a threshold in the range $9.7\pm0.2\%$. Similar to the Toric code, inspection of sub-threshold regime shows evidence of exponential suppression for the semi-topological code family. Within margin of error, the semi-topological code threshold aligns with the threshold for the Toric code using the same decoder. This is the expected behaviour, reflecting the fact that semi-topological codes become structurally similar to Toric codes (more local) as their distance is increased. The structural similarity arises because the chain-segments are mapped to toric-code-like patches by the hypergraph product, and these regions form the bulk in the limit of large $g$. Discrepancies in the threshold between the semi-topological codes and toric codes can be attributed to finite size effects.

\begin{table}[]
	\renewcommand{\arraystretch}{1.4}
	\begin{tabular}{l|c|c|c}
		Code             & BP      & BP+OSD-0 & BP+OSD-CS \\ \hline
		Toric            & N/A     & $9.2\pm0.2\%$  & $9.9\pm0.2\%$          \\
		Semi-topological & N/A     & $9.1\pm0.2\%$  & $9.7\pm0.2\%     $     \\
		Random            & $6.5\pm0.1\%$ & $6.7\pm0.1\%$  & $7.1\pm0.1 \% \ $         
	\end{tabular}
	
	\caption{Observed thresholds for numerical simulations of the BP+OSD decoder applied to Toric, semi-topological and random QLDPC codes.}
	\label{TableThresholds}
\end{table}

\textit{Random QLDPC Codes} --- Figure.~\ref{plot:qldpc} shows the results of numerical simulations of the BP+OSD decoder applied to the finite-rate family of random QLDPC codes summarised in Table.~\ref{tab:qldpc}. The code distances considered are $d=\{6,8,10\}$. In contrast to the Toric and semi-topological codes, the BP decoder alone (before any OSD post-processing) shows a crossing, pointing to a threshold in the range $6.5\pm0.1\%$. The existence of this threshold for the BP decoder can be attributed to the fact that random QLDPC codes are less structured than Toric and semi-topological codes; the repeating patterns present in stabilizer checks of topological codes lead to high densities of degenerate errors that cause BP to fail. The full BP+OSD-CS decoder applied to the random QLDPC family results in a threshold in the range $7.1\pm0.1\%$. Whilst this threshold value is only a modest improvement over BP, the real benefit of the OSD post-processing for random QLDPC codes becomes apparent in the low-error regime; at $p=0.01$, for example, the logical error rate $p_L$ for BP+OSD-CS decoder is approximately an order magnitude less than that for BP.

\section{Summary}\label{sec:summary}

Quantum LDPC codes have traditionally been studied as local topological codes or non-local random codes. In this paper we introduce semi-topological codes as a means of interpolating on the local to non-local QLDPC spectrum. Previously, the practicality of QLDPC codes has been hindered by the lack of a general purpose decoder: designing a new family of QLDPC codes would necessitate the development of a special-purpose decoding strategy \cite{fawzi2018efficient,grospellier2018numerical}. In this paper, we provide further evidence that the recently proposed BP+OSD decoder \cite{Panteleev_2019} applies to all QLDPC codes constructed via the hypergraph product, including our new family of semi-topological codes.

The methods for constructing semi-topological codes proposed in this paper allow the locality of QLDPC codes to be balanced against other factors such as code rate. The existence of a general purpose BP+OSD decoder for QLDPC codes grants quantum computer architects more freedom in the design of fault tolerant quantum computers; modifications to the structure of a QLDPC code can be made according to demands of a given device, without compromising the practicality of their decoding.

All of the simulations in this work were run under the assumption that the syndrome measurements are noiseless. In reality, syndrome extraction is performed using ancilla qubits with imperfect readout. In work currently in preparation \cite{quintavalle2020}, we study the performance of the BP+OSD decoder for higher dimensional hypergraph product codes with the single-shot property \cite{bombin2015single,vasmer2019three,campbell2019theory}, designed with in-built protection against syndrome noise.

\etc{Since our semi-topological codes contain local patches of surface code, it would be useful to determine whether other QLDPC codes can be modified to contain such patches.  For instance,  Panteleev and Kalechev \cite{Panteleev_2019} constructed a $[[1270,28,d]]$ (with unknown $d$) code that was especially competitive with surface codes.  However, this code was constructed using a generalised hypergraph product and it is unclear whether an analog of our edge-augmentation process can be applied to this more general code family.}

In this paper we have demonstrated the versatility of BP+OSD as a decoder across the spectrum of QLDPC codes that can be obtained from the hypergraph product. Beyond this, we conjecture that BP+OSD decoders will apply more generally to QLDPC codes constructed using different methods. Potential candidates for future investigation include topological fracton codes \cite{vijay2016fracton} and QLDPC codes based on high-performance classical protocols \cite{hagiwara2011spatially}.

\section*{Acknowledgements}

JR, SB and EC are supported by the QCDA project (EP/R043825/1) which has received funding from the QuantERA ERA-NET Cofund in Quantum Technologies implemented within the European Union's Horizon 2020 Programme. EC is additionally supported by the Engineering and Physical Sciences Research Council \etc{(EP/M024261/1). DW was supported by a research grant from Huawei. We thank Armanda Quintavalle for related discussions and comments throughout the project.}  The authors are grateful for the use of the following open source software packages: Software for LDPC codes \cite{radford_neal}; Scipy \cite{scipy}; Numpy \cite{numpy}; Matplotlib \cite{matplotlib}.

\section*{Software}
The code for the BP+OSD decoder used for the simulations in this paper can be downloaded from Github. \url{https://github.com/quantumgizmos/bp_osd}

\bibliographystyle{hunsrt2}
\bibliography{hgp_bp_osd_paper.bib}

\appendix

\begin{figure}
	\input{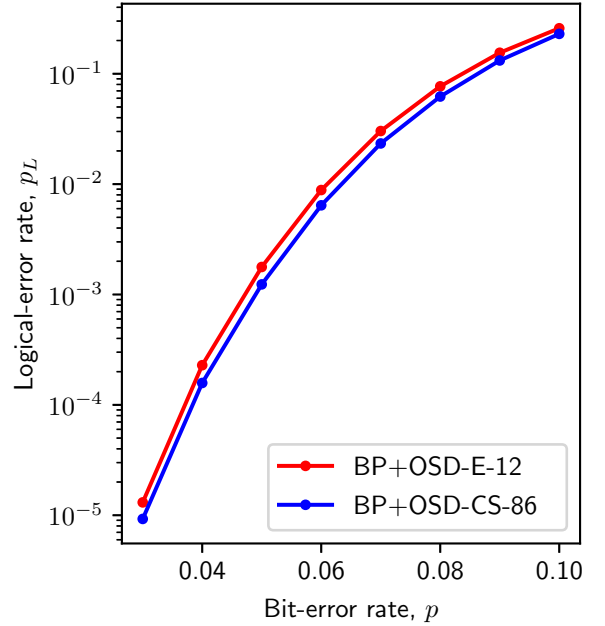}
	\caption{Comparison of the BP+OSD-E and BP+OSD-CS methods when applied to the distance $d=15$ Toric code. The $\lambda$ value for BP+OSD-E is set to $\lambda=12$, leading to a total of $4096$ inputs to the encoding operator defined in equation (\ref{eq:osd_enc}). For BP+OSD-CS, the $\lambda$ value is set to $\lambda=86$, leading to $3881$ inputs to the encoding operator.}
	\label{fig:comparison}
\end{figure}

\section{Lower bound on the increase in code distance due to edge augmentation}

\begin{theorem}
	\label{app:lower_bound}
	Consider an $[n,k,d]$ classical code with Tanner graph $G=(V,U,\Lambda)$ and let $\mu$ denote the minimum degree over all data nodes $V$.  Let $G^{\star g}$ be the Tanner graph resulting from augmenting each edge of $G$ with $g$ data nodes and $g$ parity nodes.  It follows that $G^{\star g}$ corresponds to an $[ n +g |\Lambda| , k , d'  ]$ code with $d' \geq (1+g \mu)d$.
\end{theorem}

\begin{proof}  	For the augmented graph $G^{\star g}$, we divide the data qubits $V \cup V^{g}$ into two disjoint subsets: the parent data nodes $V$ and the augmented data nodes $V^{g}$. We let $A$ denote a subset of data qubits $A \subseteq V \cup V^{g}$ that corresponds to a codeword of the classical code, which is the case if and only if every check node in $G^{\star g}$ has an even number of graph neighbours in the set $A$.	 Furthermore, because the augmented data nodes $V^{g}$ are all degree two, for each graph chain segment either all the data nodes are in $A$ or none of them are.  Furthermore, for every parent data node $a \in A \cap V$, it follows that every whole graph chain segment welded to $a$ must be in $A$.  Using  $\#\mathrm{chains}(A)$ to denote the number of graph chain segments present in $A$, we have that
\begin{equation}
	|A| =  | A \cap V  |  + g  \#\mathrm{chains}(A)\rm.
\end{equation}		
Recall that each graph chain segment welds to one parent data node and one parent check node.  Therefore, we can count the number of graph chain segments as follows $\#\mathrm{chains}(A)= \sum_{a \in A \cap V} \mathrm{deg}(a)$.  In the graph $G^{\star g}$, the parent data nodes have the same degree as they did in the original graph $G$ and by assumption this is lower bounded by $\mu$.  Therefore, we have $\#\mathrm{chains}(A) \geq \mu |A \cap V|   $ and so
\begin{equation}
	\label{Acounting}
	|A| \geq (1+ g \mu) | A \cap V  |.
\end{equation}	
Next, we observe that $A$ can only be a codeword with respect to graph $G^{\star g}$ if $A \cap V$ is a codeword with respect to graph $G$, which entails that 
\begin{equation} 
	\label{AVlower}
	| A \cap V  | \geq d. 
\end{equation}
Let us break this observation down into steps.  Assume to the contary that 	$| A \cap V  | < d$, so that with respect to graph $G$ there is a parent check node $c \in U$ such that it has an odd number of neighbours in $A \cap V$.  Furthermore, there will be an odd number of edges connecting $c$ to $A \cap V$ in graph $G$.   Each one of these edges maps to a graph chain segment in $A \cap V$ in the augmented graph, each of which welds to check node $c$.  Therefore, check node $c$ also has an odd number of neighbours with respect to graph $G^{\star g}$.  This is impossible when $A$ is a codeword in graph $G^{\star g}$, so we must have that Eq.~\eqref{AVlower} holds. Combining Eq.~\eqref{Acounting} and Eq.~\eqref{AVlower}, gives $|A| \geq (1+ g \mu) d$ for any codeword $A$ in $G^{\star g}$, so this gives a lower bound on $d'$.

\end{proof}

\section{Comparison of BP+OSD greedy search algorithms}\label{app:comparison}

The greedy search stage of a BP+OSD decoder involves testing different inputs to the OSD encoding operator given by equation (\ref{eq:osd_enc}), with the aim of finding solutions that improve upon OSD-0. In \cite{Panteleev_2019}, Panteleev and Kalachev use an exhaustive search strategy which we refer to as BP+OSD-E. For the simulations in this paper, we adopt the combination sweep method, referred to as BP+OSD-CS. For an OSD solution of the form, $\textbf{e}_{OSD}=(\textbf{e}_{[S]},\textbf{e}_{[T]})$, the first step is to order the bits in $\textbf{e}_{[T]}$ according to the soft-decisions from BP. The greedy search for the two methods then proceeds as follows:
\begin{enumerate}
	\item BP+OSD-E: All permutations of most-probable $\lambda$ bits in $\textbf{e}_{[T]}$ are searched over. In total, $2^\lambda$ search terms are considered.
	\item BP+OSD-CS: All weight one and two permutations in the first $\lambda$ most probable bits of $\textbf{e}_{[T]}$ are searched over. The total number of configurations considered is equal to $k'+{\lambda \choose 2}$, where $k'=n-\text{\sc rank}(H)$ is the length of the $\mathbf{e}_{[T]}$ vector.
	  
\end{enumerate}
Figure.~\ref{fig:comparison} compares decoding performance for the above OSD methods when applied to the $d=15$ Toric code. For the exhaustive method (BP+OSD-E), $\lambda$ is set to $\lambda=12$, giving a total of $4096$ search terms. For the combination sweep method (BP+OSD-CS), $\lambda$ is set to $\lambda=86$, giving a total of $3881$ search terms. Figure.~\ref{fig:comparison} shows that, despite considering fewer search terms, the BP+OSD-CS method improves decoding performance compared to BP+OSD-E.

\section{Min-sum Belief Propagation}\label{app:bp}

We gave only a short summary of BP in the paper; we provide further details here to avoid the ambiguity often found in the literature, where BP is formulated for different applications and with many variations. We use a variant of the ``min-sum'' algorithm, using log-likelihood ratios for probabilities and the incorporation of variable scaling to prevent runaway values.
 
Belief Propagation calculates marginal probabilities over graphical probabilistic models, a form of statistical inference, and is widely applied to the decoding of classical error-correcting codes. In the quantum domain the decoding task differs slightly in that, rather than trying to infer the original codeword from the received message, we are given the syndrome indicating whether a given ``parity check'' failed and must infer a recovery operator; we must also cope with quantum degeneracy. Despite these differences, the task of quantum error correction can be reformulated as a classical syndrome-based decoding problem. Unfortunately,  syndrome-based decoding is not common in the classical decoding literature and there are few good references on the topic.

A more significant difference when applying BP to quantum codes is that all CSS and non-CSS QLDPC codes have factor graphs of girth four; originally BP was designed to work on acyclic graphs, but these factor graphs contain short cycles. Whilst this violates the invariants of the algorithm and hence its proof of its correctness, empirically BP has been found to perform surprisingly well on cyclic graphs. Although it may sometimes fail to converge to a feasible solution, we can detect this by checking that its output satisfies the syndrome equation.

\subsection*{Formulating QEC Decoding for BP}
A QEC factor graph has data nodes representing each bit in the error string, which we denote $v_j$. It has one ``check'' or ``parity'' node for each syndrome measurement, which we denote $u_i$. The graph is described by the parity check matrix $H$ (whether it concerns X or Z errors alone, or both, is immaterial; the methodology is the same). A one at position $(i, j)$ in $H$ indicates that parity node $u_i$ has an edge directly connecting it to data node $v_j$.

There are two forms of prior information we must incorporate into the graph: the error rate of the channel, $p$, and the syndrome $s$. The error rate is incorporated as a hidden input to the data nodes. The syndrome measurement is implicitly present in the graph via calculations made at the parity nodes.

BP is conceptualised as a message passing algorithm. We denote a message from a parity to a data node $m_{u_i \rightarrow v_j}$ and from a data node to a parity node as $m_{v_j \rightarrow u_i}$. As we possess only the syndrome, and not the received codeword, the factor graph for QEC is slightly different from the standard graph found in classical decoding --- but it is indeed equivalent to the (rarely discussed) syndrome-based classical decoding.

Our task is as follows: given the syndrome $s$ and the structure represented by the factor graph, what is the most likely value of each bit in the error string?

\subsection*{Algorithm Description}
\begin{algorithm}[H]
\begin{algorithmic}[1]
\Statex
\Function{BeliefProp}{$H$, $\textbf{s}$, $p$}  
\Statex
\State $\triangleright$ Channel LLR
\Let{$p_l$}{\mbox{log}((1 - $p$)/$p)$} 
\Statex
\State $\triangleright$ (1) Initialisation
\For{$(v_j, u_i) \in H$} 
\Let{$m_{v_j \rightarrow u_i}$}{$p_l$} 
\EndFor
\Statex
\For{$iter \gets 1 \textrm{ to }max$}
\Statex
\State $\triangleright$ Scaling Factor
\Let{$\alpha$}{$1 - 2^{-iter}$} 
\Statex
\State $\triangleright$ (2) Parity to Data Msgs
\For{$(u_i, v_j) \in H$} 
\Let{$w$}{$\mbox{min}_{\sim v_j \in V(u_i)} \{ |m_{v_j' \rightarrow u_i}| \}$}
\State $m_{u_i \rightarrow v_j} = -1^{s_i} \alpha ( \prod_{\sim v_j \in V(u_i) } sign ( m_{v_j' \rightarrow u_i}) ) w$
\EndFor   
\Statex
\State $\triangleright$ (3) Data to Parity Msgs 
\For{$(v_j, u_i) \in H$} 
\Let{$m_{v_j \rightarrow u_i}$}{$m + \sum_{\sim u_i \in U(v_j)} m_{u_i' \to v_j}$}
\EndFor
\Statex
\State $\triangleright$ Hard Decision
\For{$(v_j, u_i) \in H$} 
\Let{$P_1(e_j)$}{$p_j + \sum_{u_i' \in U(v_j)} m_{u_i' \to v_j}$}
\Let{$\mathbf{e}_{BP}^{j}$}{$- sign(P_1(e_j))$}
\EndFor  			   		
\Statex
\State $\triangleright$ (4) Termination Check    						
\If{$H \cdot \mathbf{e}_{BP} = \textbf{s}$} 
\State \Return{True, $\mathbf{e}_{BP}$, $P_1$}
\EndIf
\EndFor
\Statex
\State $\triangleright$ Failed to Converge
\State \Return{False, $\mathbf{e}_{BP}$, $P_1$} 
\EndFunction
\end{algorithmic}
\caption{Pseudocode for Belief Propagation using log likelihood ratios, the minsum product algorithm, and a scaling factor. Log likelihood ratios and the minsum algorithm (the use of $w$ in Line 13) make the computation more efficient and avoid the numerical instability of other implementations.}
\label{alg:bp}
\end{algorithm}

The pseudocode for our implementation is given in Algorithm \ref{alg:bp}, and consists of four sequential steps:

\begin{enumerate}

	\item Initialisation
	
	Messages are sent from data nodes to parity nodes giving the a priori probability of that bit in the error string being a one, i.e. the LLR (log likehood ratio) of the channel error rate $p$, which we denote $p_l$ in its log likelihood form:

	\begin{equation}
		p_l \triangleq log( (1-p) / p )
	\end{equation}

	\item Parity nodes to data nodes
	
	Messages are sent from parity nodes to data nodes containing the marginal probability of an error at the destination data node. However, we implement several optimisations that somewhat complicate the calculation of this message. Denoting the neighbouring data nodes of a given parity node $u_i$ as $V(u_i)$, the messages sent are:
	
\begin{align}
	m_{u_i \rightarrow v_j} &= \etc{-1^{s_i}} \alpha \\
	&\cdot \Bigg( \prod_{v_j' \in V(u_i) \setminus v_j} \mbox{sign} ( m_{v_j' \rightarrow u_i}) \Bigg) \\ 	
	&\cdot \mbox{min}_{v_j' \in V(u_i) \setminus v_j} \{ |m_{v_j' \rightarrow u_i}| \}
\end{align}

	The set minus notation in the subscripts indicates that this is a marginal distribution, i.e. we consider only the probabilities from other data nodes when calculating the marginal for this bit. The sign function and the first exponential $(-1)^{s_i}$ are used to incorporate the syndrome\etc{, with $s_i$ being the $i^{\mathrm{th}}$ bit of the syndrome}. In other words: ``consider all configurations of connected error bits, and increase the probability of the implied value for this bit compatible with the observed syndrome.''  The first factor is an XOR operation that establishes the sign of this probability, i.e. whether $u_i$ is implied to be a one or a zero, based on the decision represented by the messages sent by other data bits. The second factor describes the magnitude of the probability, is based on the notion that the `cheapest' way that this value of $u_i$ could be incorrect is if one of the other bits was flipped. For a full explanation, see \cite{kschischang2001factor}.
	
We also include $alpha$, a scaling factor as outlined in \cite{emran2014simplified}. The scaling factor $\alpha$ is set according to the current iteration $iter$, where the first iteration is numbered $iter=1$:
		
\begin{equation}
\alpha = 1 - 2^{-iter}
\end{equation}

	\item Data nodes to parity nodes
	
	Next, messages are sent from data nodes to parity nodes giving the probability ratio for that bit in the error string, calculated by summing the inbound marginals and taking into account the error rate for the channel, omitting normalisation for efficiency:
	
\begin{equation}
m_{v_j \rightarrow u_i} = p_l + \sum_{u_i' \in U(v_j) \setminus u_i} m_{u_i' \to v_j}
\end{equation}

\etc{Where we have denote the neighbouring data nodes of a given check node $v_j$ as $U(v_j)$.}

	\item Termination check. If the factor graph is a tree, we can always terminate after a single iteration of the algorithm. If it is cyclic (as in QEC), then we will terminate on success or else when a given number of iterations are complete. We first calculate a ``hard decision'' of the most likely error string, by selecting the most likely configuration via the bitwise marginals we have calculated:
	
\begin{equation}
P_1(e_j) = p_l + \sum_{u_i' \in U(v_j)} m_{u_i' \to v_j}
\end{equation}
	
	We then select the most likely error string $\tilde E$ given these bitwise probabilities, and calculate the expected syndrome:
	
	\begin{equation}
		\textbf{s} = H \cdot \mathbf{e}
	\end{equation}
	
	We terminate if $\mathbf{s}$ matches the measured syndrome, or if we have reached a preset maximum number of iterations (often equal to the block length). Otherwise, we return to Step 2, sending `parity nodes to data nodes' messages.
	
\end{enumerate}

The outputs of BP are both the soft and hard decisions; the former are used by OSD if BP has failed to converge, i.e. the hard decision does not satisfy the syndrome equation. The soft decision is a bit-wise estimate of the probability an error occurred, which OSD uses to bias its search for an error string.

\end{document}